\RequirePackage{rotating}
\documentclass[12pt,reqno]{amsart}
\usepackage{amsaddr}

\usepackage[utf8]{inputenc}
\usepackage{xcolor}
\usepackage{acro}
\usepackage{amsmath}                        % assumes amsmath package installed
\usepackage{amsfonts}                       % assumes amsmath package installed
\usepackage{amssymb}                        % assumes amsmath package installed
\usepackage{amsthm}
\usepackage{graphicx}                       % for pdf, bitmapped graphics files
\usepackage{enumitem}
\usepackage[binary-units]{siunitx}
\usepackage{subcaption}
\usepackage{optidef}
\usepackage[hidelinks]{hyperref}
\usepackage{float}
\usepackage{subcaption}
\usepackage{textcomp}
\usepackage{graphicx}
\usepackage{fullpage}

\theoremstyle{definition}

\theoremstyle{remark}

\newcommand{\transposed}{^\mathsf{T}}

\newcommand{\ud}{u_{\dot{\delta}}}
\newcommand{\uv}{u_{\dot{v}}}

\newcolumntype{L}[1]{>{\raggedright\let\newline\\\arraybackslash\hspace{0pt}}m{#1}}
\newcolumntype{C}[1]{>{\centering\let\newline\\\arraybackslash\hspace{0pt}}m{#1}}
\newcolumntype{R}[1]{>{\raggedleft\let\newline\\\arraybackslash\hspace{0pt}}m{#1}}

\definecolor{marine}{RGB}{0,128,128}
\definecolor{wine}{RGB}{128,0,0}
\definecolor{oceanus}{RGB}{0,0,128}
\DeclareAcronym{cg}{
    short = CG ,
    long  = center of gravity
}
\DeclareAcronym{cmpc}{
    short = CMPC ,
    long  = centralized model predictive control
}
\DeclareAcronym{cpmlab}{
    short = CPM Lab,
    long  = Cyber-Physical Mobility Lab
}
\DeclareAcronym{ctg}{
    short = CTG ,
    long  = cost-to-go
}
\DeclareAcronym{ctc}{
    short = CTC ,
    long  = cost-to-come
}
\DeclareAcronym{dmpc}{
    short = DMPC ,
    long  = distributed model predictive control
}
\DeclareAcronym{dag}{
    short = DAG ,
    long  = directed acyclic graph
}
\DeclareAcronym{fsa}{
    short = FSA ,
    long  = finite state automaton,
    short-indefinite = an,
}
\DeclareAcronym{lwa}{
    short = LWA* ,
    long  = lazy weighted A*,
    short-indefinite = an,
}
\DeclareAcronym{lsp}{
    short = LazySP ,
    long  = lazy shortest path,
    short-indefinite = an,
}
\DeclareAcronym{lra}{
    short = LRA* ,
    long  = lazy receding horizon A*,
    short-indefinite = an,
}
\DeclareAcronym{ma}{
    short = MA ,
    long  = maneuver automaton,
    short-indefinite = an,
}
\DeclareAcronym{mip}{
    short = MIP,
    long  = mixed integer programming,
    short-indefinite = an,
}
\DeclareAcronym{mld}{
    short = MLD ,
    long  = mixed logical dynamical,
    short-indefinite = an,
}
\DeclareAcronym{mpa}{
    short = MPA ,
    long  = motion primitive automaton,
    short-indefinite = an,
}
\DeclareAcronym{mpc}{
    short = MPC ,
    long  = model predictive control,
    short-indefinite = an,
}
\DeclareAcronym{ncs}{
    short = NCS ,
    long  = networked control system,
    short-indefinite = an,
}
\DeclareAcronym{ocp}{
    short = OCP ,
    long  = optimal control problem,
}
\DeclareAcronym{mocp}{
    short = MOCP ,
    long  = multiobjective optimal control problem,
}
\DeclareAcronym{pbnc}{
    short = NCPB ,
    long  = priority-based\, non-cooperative
}
\DeclareAcronym{ode}{
    short = ODE ,
    long  = ordinary differential equation
}
\DeclareAcronym{qp}{
    short = QP,
    long  = quadratic programming,
}
\DeclareAcronym{rhgs}{
    short = RHGS ,
    long  = receding horizon graph search,
    short-indefinite = an,
}
\DeclareAcronym{rhc}{
    short = RHC ,
    long  = receding horizon control,
    short-indefinite = an,
}
\DeclareAcronym{rrt}{
    short = RRT ,
    long  = rapidly-exploring random tree,
    short-indefinite = an,
}
\DeclareAcronym{sgs}{
    short = SGS ,
    long  = state-of-the-art graph search,
    short-indefinite = an,
}
\DeclareAcronym{tsp}{
    short = TSP ,
    long  = traveling salesman problem,
}
\DeclareAcronym{uav}{
    short = UAV ,
    long  = unmanned aerial vehicle
}

\DeclareAcronym{mp}{
    short = MP ,
    long  = motion primitives
}

\DeclareAcronym{kst}{
    short = KST ,
    long  = kinematic single-track
}

\DeclareAcronym{ua}{
    short = UA ,
    long  = universal automaton
}

\begin{document}
	
	\title{Optimization-based motion primitive automata for autonomous driving}
	
	\author{Matheus V. A. Pedrosa$\hspace{0.5ex}^1$, Patrick Scheffe$\hspace{0.5ex}^2$, Bassam Alrifaee$\hspace{0.5ex}^2$ \and Kathrin Fla$\mathcal{B}$kamp$\hspace{0.5ex}^1$}
	
	\address{$^1$ Chair of Systems Modeling and Simulation, Systems Engineering, Saarland University, Germany}\email{\{matheus.pedrosa, kathrin.flasskamp\}@uni-saarland.de}
	
	\address{$^2$ Chair for Embedded Software, RWTH Aachen University, Germany}\email{\{scheffe, alrifaee\}@embedded.rwth-aachen.de}
	
	\thanks{This research is supported by the Deutsche Forschungsgemeinschaft (German Research Foundation) within the Priority Program SPP 1835 ``Cooperative Interacting Automobiles" (grant number: KO 1430/17-1).}
		
	\begin{abstract}
		Trajectory planning for autonomous cars can be addressed by primitive-based methods, which encode nonlinear dynamical system behavior into automata. In this paper, we focus on optimal trajectory planning. Since, typically, multiple criteria have to be taken into account, multiobjective optimization problems have to be solved. For the resulting Pareto-optimal motion primitives, we introduce a universal automaton, which can be reduced or reconfigured according to prioritized criteria during planning. We evaluate a corresponding multi-vehicle planning scenario with both simulations and laboratory experiments.
	\end{abstract}
	
	\maketitle

	\section{Introduction}
	Optimization-based control techniques play a crucial role in autonomous driving.
	Focusing on trajectory generation, we apply a graph-based planning method based on motion primitives, where optimization criteria can be considered at all stages of the method, i.e., the construction of the primitives and the motion planning.
	For computing motion primitives, we formulate multiobjective optimal control problems. The special structure-preserving primitives, named \textit{trims}, are equipped with unit costs, which can be evaluated for selecting a set of representative trims.
	While classical, continuous-time optimal control does not easily allow for switching objectives, the graph-based approach introduces dynamic waypoints via the trim primitives, at which prioritization of criteria, defined by the cost and heuristic functions, can be switched \cite{Fla2013}.
	However, the multi-objective design choices lead to different automata (graphs), and including Pareto-optimal motion primitives strongly influences the size of each automaton.
	Thus, we evaluate and compare the performance of different subgraphs of a general-purpose automata, named \textit{universal automata}, in simulation and experiments in the \ac{cpmlab}, an open-source platform for networked and autonomous vehicles \cite{klock-cyber-lab-2021}.

	\section{Preliminaries}
	
	Motion planning with symmetry-exploiting primitives has been proposed by Frazzoli et al. in \cite{frazzoli2005} as a general concept.
	For shortness, we summarize its application to vehicle planning in the following.

	We consider the \ac{kst} model for the vehicles with the state vector
	$x = \begin{bmatrix} s_x & s_y & \psi & v & \delta \end{bmatrix}\transposed \in \mathcal{M} \subset \mathbb{R}^5$,
	and the input vector:
	$u = \begin{bmatrix} \uv & \ud \end{bmatrix}\transposed \in \mathbb{R}^2$,
	where $s_x$ and $s_y$ are the positions of the \ac{cg}, $\psi$ is the vehicle orientation, $v$ is the velocity, $\delta$ is the steering angle, $\mathcal{M}$ is the 5-dim state manifold, $\uv$ is the longitudinal acceleration, and $\ud$ is the velocity of the steering angle.
	The dynamics are given by:
	{\small
		\begin{equation} \label{eq:kst_model}
			\dot{x}(t) = f \big( x(t), u(t) \big) := 
			\begin{bmatrix}
				v(t) \cdot \cos(\psi(t) + \beta(t)) \\
				v(t) \cdot \sin(\psi(t) + \beta(t)) \\
				\frac{v(t)}{L} \cdot \tan(\delta(t)) \cos(\beta(t)) \\
				\uv(t) \\
				\ud(t) \\
			\end{bmatrix}
		\end{equation}
	}with
	$\beta(t) = \tan^{-1}\left(\frac{l_r}{L} \tan(\delta(t)) \right)$,
	where $L$ is the wheelbase length and $l_r$ is the length from the rear axle to the \ac{cg} \cite{commonroad-Althoff2017a}.

	For any given control signal $u$ on $[0,T]$, motions are invariant w.r.t.\ combined translations and rotations, i.e.,\ let $ x(t) = \varphi_u(x_0,t)$ denote the trajectory starting at time $t=0$ in $x_0$, 
	then it holds
	$\varphi_u(\Psi_g(x_0),t) = \Psi_g(\varphi_u(x_0,t))$ for all $t\in [0,T]$ with $g=\begin{pmatrix} \Delta s_x & \Delta s_y & \Delta \psi \end{pmatrix}$ and
	{\small
		$$ \Psi_g(x) = \begin{bmatrix} 
			\begin{bmatrix}
				\cos(\Delta \psi) & - \sin(\Delta \psi) & 0 \\
				\sin(\Delta \psi) & \cos(\Delta \psi) & 0 \\
				0 & 0 & 1 
			\end{bmatrix} \begin{bmatrix} s_x\\ s_y \\ \psi \end{bmatrix} + \begin{bmatrix} \Delta s_x \\ \Delta s_y \\ \Delta \psi \end{bmatrix} ; & 
			\begin{bmatrix} v & \delta \end{bmatrix}^T
		\end{bmatrix},          
		$$
	}see e.g.\ \cite{scheffe2022ieee, learning-pedrosa2022} for a proof. The equality above describe the symmetry property.
	In turn, all trajectories being equal up to the shift by $\Psi_g$ for all $g\in \mathcal{G}$, i.e.\ the symmetry group, are summarized in the equivalence class called \emph{motion primitive}.

	A special class of motion primitives are those with fixed -- i.e.\ trimmed -- control, called \emph{trim primitives}, which possess an analytic expression thanks to symmetry, cf. \cite{frazzoli2005}.
	For the \ac{kst} model, any motion with constant longitudinal velocity and fixed steering angle, i.e.\ $\uv(t) =
	\ud(t) = 0 $ satisfies this condition. Omitting technical details, these motions belong to either "driving straight with constant velocities" or "driving on arcs of circles with constant velocities".
	
	Motion primitives can be \emph{concatenated}, if symmetry shifts $\Psi_g$ can be used to map a successor's starting point to the predecessors ending point (again, cf.~\cite{frazzoli2005} or the authors' previous works for details).
	For switching between one trim primitive to another, a special connecting primitive is needed; e.g., it can accelerate, decelerate or adjust the steering angle. This kind of primitive is named \emph{maneuver}.
	Thus, maneuvers typically require highly controlled motion, so that optimal control is a suitable tool to compute maneuvers (see Sect. \ref{sec:maneuver_design}).
	Sequences of motion primitives which form valid trajectories of Eq.~\eqref{eq:kst_model} alternate between trim primitives and maneuvers.
	
	Finally, the idea of motion planning with primitives is to restrict to a \emph{finite set of motion primitives} and solve a given planning problem by finding the best sequence within this set.
	Given the distinction into trim primitives and maneuvers, it is convenient to represent the finite set of motion primitives by a graph \cite{frazzoli2005}.
	The trims are represented by the vertices of the graph, while the maneuvers, by the edges.
	For a fully state/control-quantized and time-discretized setting, also the duration of trims is defined to be fixed.

	\section{Optimal Motion Primitive Automaton Design}
	\label{sec:maneuver_design}

	Selection strategies for trims to be considered in the motion primitive graph that can be found in the literature are, e.g., an arbitrary choice according to the expected operating points \cite{Flakamp2019SymmetryAM, pedrosa2021} or detecting the most representative trims from a dataset \cite{learning-pedrosa2022}.
	Complementarily, we focus on the optimization criteria here.
	Let $J(T,x,u) = \int_0^T \ell(x(t),u(t)) \, dt$ be a cost function with running cost $\ell(\cdot,\cdot)$, i.e.\ any criterion depending on a state-control pair, which is also invariant w.r.t.\ $\Psi_g$, the system dynamics' symmetry.
	Then, for state-control signals $(\tilde{x},\tilde{u})$  being trim primitives, it holds that
	$$J(T,\tilde{x},\tilde{u}) = T \cdot  \ell(\tilde{x}(0),\tilde{u}(0))= T \cdot  \ell(\tilde{x}(0),0). $$
	While the first equality is due to trim definition ($\tilde{u}(t) = \text{const.}$, in particular) and invariance (see \cite[Prop. 4]{Flakamp2019SymmetryAM}), the second simplification holds for the \ac{kst} model, since any trim satisfies $\tilde{u} \equiv 0$. $\ell(\tilde{x}(0),\tilde{u}(0))$ is called \emph{unit cost} in \cite{frazzoli2005}.

\begin{figure}
		\centering
		\includegraphics[width=\textwidth]{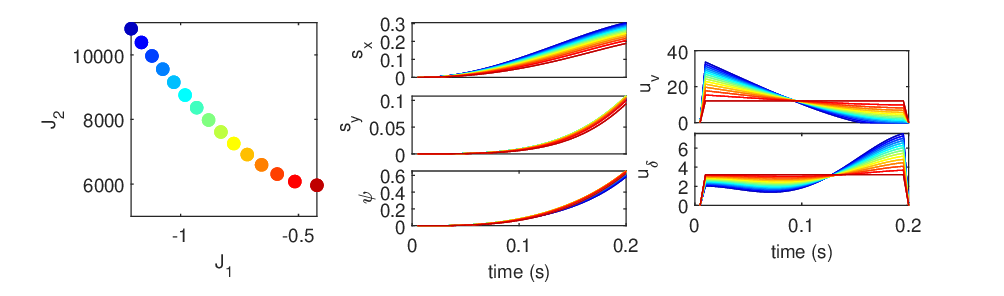}
		\caption{Example of a Pareto set with respective positions, orientations, and inputs for a maneuver going from a trim with zero speed and steering angle to $v=\SI{2.3}{\meter\per\second}$ and $\delta = \SI{0.62}{\radian}$. It was considered the \ac{kst} model with $J_1= -\int_0^T (||s_x||_2^2 + ||s_y||_2^2) dt$ and $J_2=\int_0^T ||u||_2^2dt$, for a fixed duration $T=\SI{0.2}{\second}$, and the parameters used in the evaluation.}
		\label{fig:pareto_set}
	\end{figure}

	To obtain maneuvers that are optimal with respect to multiple criteria $J_1, \dots, J_K$, $K>1$, the following {\ac{mocp}} can be solved:
	\begin{mini!}
		{T,x,u}{\begin{bmatrix} J_1(T,x,u) & \dots & J_K(T,x,u) \end{bmatrix}^{\transposed} \label{optim:objectivefcn_man}}{\label{optim_problem_man}}{}
		\addConstraint{\dot{x}(t)}{=f(x(t),u(t)),}{\hspace{1ex} 0 < t \leq T, \hspace{1ex} \text{cf.\ Eq.~\eqref{eq:kst_model}}}
		\addConstraint{0}{= r(x(0), x(T))}{}
		\addConstraint{0}{\geq g(x(t),u(t)),}{\hspace{1ex} 0 < t \leq T,}
	\end{mini!}
	with boundary constraints $r(x(0), x(T))$ assuring the connection of trims, and~$g(\cdot)$ for any type of state/input constraints in between.\\
	In autonomous driving, different kinds of objectives may play a role, e.g.:\\[-.8cm]
	\begin{itemize}
		\item Travel time: not only in transportation of goods, minimizing travel time corresponds to higher profits.
		\item Effort: saving energy is of critical importance when the fuel or the battery is close to empty, or for reducing monetary costs.
		\item Safety: emergency maneuvers can be needed for critical situations \cite{naumann2018_coincar}. 
		\item Passengers' comfort: typically modeled via acceleration minimization or by avoiding sickness-proven eigenmodes in multi-body systems for car-human models. It is assumed that autonomous driving should improve safety without losing too much comfort or utility \cite{naumann2018_coincar}.
	\end{itemize}
	In \autoref{fig:pareto_set}, we show an example of a Pareto set for maneuvers going from a trim with $v=\SI{0}{\meter\per\second}$ and $\delta = \SI{0}{\radian}$ to $v=\SI{2.3}{\meter\per\second}$ and $\delta = \SI{0.62}{\radian}$, but computed with different cost functionals. It was considered the same constraints for the model-scale vehicle used in the evaluation \cite{commonroad-Althoff2017a}. The duration was a fixed parameter, as a requirement to be explained in the next section. The Pareto set shows a trade-off between travel time in $J_1$ and effort/comfort in $J_2$. Note that the maneuvers in blueish colors show a longer traveled distance with high control inputs in opposition to soft inputs for a smaller distance in the reddish points. 
	
	Following the strategies for obtaining primitives mentioned above, we can define a general-purpose automaton named \emph{\ac{ua}} (see \autoref{fig:universal_automaton}). It can include many trim primitives, and, in particular, the Pareto-optimal maneuvers for connecting pairs of those trims \cite{Fla2013}.
	Along the way, the \ac{ua} can reconfigure its size and operation mode by enabling/disabling some trims and/or maneuvers. Firstly, it can restrict the trims to respect, for example, the allowed speed at a certain road. Second, choosing the objective $J_k$ ($1\leq k\leq K$), which is currently of prior importance at the time of the planning task.
	Then, planning is performed on the subgraph only consisting of the maneuvers optimal to $J_k$.
	
	For the planning algorithm via graph search, the cost and heuristic functions could follow the reconfiguration of the \ac{ua} and be aligned to the choice of $J_k$. In this work, these functions will be kept constant as given in \cite{scheffe2022ieee}, since our focus is on the automaton design, rather than on the parameters of the planning method.

	\section{Evaluation and Results}
\begin{figure}
		\centering
		\begin{minipage}{.49\textwidth}
			\centering
			\includegraphics[trim={1.3cm 0.3cm 0 0},clip, scale=1]{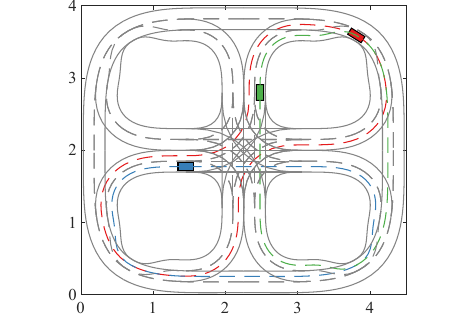}
			\captionof{figure}{Test scenario for three cars.}
			\label{fig:test_scenario}
		\end{minipage}% 
		\begin{minipage}{.49\textwidth}
			\centering
			\includegraphics[trim={3cm 0.77cm 0 0.5cm}, scale=0.75]{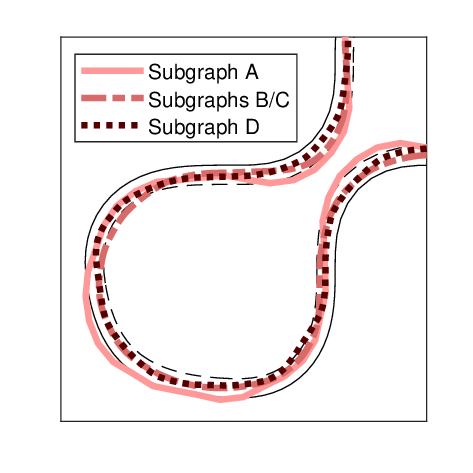}
			\captionof{figure}{Simulation of the red vehicle.}
			\label{fig:trajectories}
		\end{minipage}
	\end{figure}

	We evaluate the different automata in simulations and experimentally. In a simulator of the \ac{cpmlab}, we test three vehicles driving in collision route as illustrated in \autoref{fig:test_scenario}. The car depicted in red drives a figure-eight path, while the ones depicted in blue and green drive a circular path on the lower half and the right half of the map, respectively. In the \ac{cpmlab}, we evaluate the trajectory just for the car depicted in red.
	The simulations run on a laptop with $\SI{1.9}{\giga\hertz}$ Intel\textsuperscript{\textregistered} Core{\texttrademark} i7 CPU and $\SI{16}{\giga\byte}$ RAM.

	The \ac{cpmlab} is a 1:18-scale, open-source platform for up to 20 networked and autonomous vehicles, the \textmu{}Cars, for applications and tests on networked decision-making, trajectory planning, and control \cite{klock-cyber-lab-2021}. It runs trajectory planning algorithms on an AMD Ryzen\texttrademark \hspace{1ex}5 5600X 6-core $\SI{3.7}{\giga\hertz}$ CPU and
	$\SI{32}{\giga\byte}$ of RAM machine. It replicates a wide variety of common traffic scenarios with a 16-lane urban intersection, a highway, highway on-ramps, and highway off-ramps.
	
	The used planning algorithm is an extension of the \ac{rhgs} algorithm, presented in \cite{scheffe2022ieee}, using prioritized \ac{dmpc} \cite{scheffe2022feasibility, alrifaee2016}. The \ac{rhgs} merges the motion primitive graph search with a receding horizon framework while ensuring recursive feasibility. We consider a prediction horizon of $8$ time steps and a time step duration of $\SI{0.2}{\second}$. 
	Therefore, all maneuvers are computed for a fixed duration of $\SI{0.2}{\second}$.

	\begin{figure}
		\centering
		\includegraphics[scale=0.85]{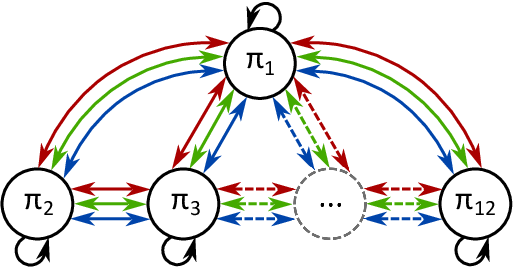}
		\caption{The universal automaton used in the evaluation for the trims given in \autoref{tab:trims}. Blue maneuvers were computed for $J_1$, red ones were computed for $J_2$, and green ones, for $J_3$. Bidirectional arrows represent two maneuvers, one in each direction.}
		\label{fig:universal_automaton}
	\end{figure}
	
	The \ac{ua} design is shown in \autoref{fig:universal_automaton} with trims given in \autoref{tab:trims} and
	maneuvers computed optimization-based w.r.t.
	$J_1= -\int_0^T (||s_x||_2^2 + ||s_y||_2^2) dt$, $J_2=\int_0^T ||u||_2^2dt$ and $J_3 = 0.5J_1 + 0.5J_2$ (see \autoref{fig:pareto_set} for a compromise between these costs), using the \ac{kst} model, initial states with zero values and respecting the constraints imposed by the \textmu{}Cars \cite{scheffe_muCars}.
	We evaluate the \ac{ua} in four configurations:
	subgraph A (B, C, respectively) has all trims and optimal maneuvers w.r.t.\ $J_1$ ($J_2$, $J_3$, resp.); subgraph D contains trims $\pi_i$, $i=1,2,4,6,7,8,10,12$, and optimal maneuvers w.r.t.\ $J_3$.
	When positioned side by side, the subgraphs A, B, and C compare different Pareto-optimal maneuvers, while C and D compare subgraphs of different sizes.
	
	\autoref{tab:results} shows the numerical results for $150$ steps, totaling $\SI{30}{\second}$. 
	We evaluate the performance through the deviation from the aimed center of the lane.
	Also, the time to complete one lap is of interest. These two evaluation parameters are from the red vehicle of \autoref{fig:test_scenario}. The trajectories for this vehicle in the simulation can be seen in \autoref{fig:trajectories}, in which the results for the subgraphs B and C are essentially similar.
	The necessary computational burden is given by the number of expanded vertices during the search. Finally, the max. computation time 
	indicates
	the complexity of the respective subgraph.

	Subgraph A shows the poorest performance.
	Since these maneuvers have the largest displacements, they can hardly track the center lane.
	Due to this fact, planning in real-time failed for subgraph A.
	Subgraphs B and C give different results in the \ac{cpmlab}, which indicates that a single change in the selected Pareto-optimal maneuvers of the \ac{ua} influences the graph search results.
	Subgraph D presents a reduced number of expanded vertices when compared to C, which reduces its computation times.
	At the cost of a slightly larger max. computation time, a solution with less deviation to the center lane is found in subgraph D, despite providing fewer trims and maneuvers here.

	\begin{table}
		\caption{Trim primitives of the universal automaton.}
		\begin{tabular}{l|llllllllllll}
				Trim ID & $\pi_1$ & $\pi_2$ & $\pi_3$ & $\pi_4$ & $\pi_5$ & $\pi_6$ & $\pi_7$ & $\pi_8$ & $\pi_9$ & $\pi_{10}$ & $\pi_{11}$ & $\pi_{12}$ \\ \hline
				$v$ [\si{\meter\per\second}] & 0&0.4&0.5&0.6&0.7&0.8&0.8&0.8&0.7&0.6&0.5&0.4 \\
				$\delta$ [\si{\radian}] & 0 & -0.60 & -0.48 & -0.36 & -0.24&-0.12&0&0.12&0.24&0.36&0.48&0.60 \\
		\end{tabular}
		\label{tab:trims}
	\end{table}
	
	\begin{sidewaystable}
		\centering
		\caption{Results, where mean values are followed by standard deviations in parentheses.}
			\begin{tabular}{l|l|l@{\hspace{1\tabcolsep}}l|l@{\hspace{1\tabcolsep}}l|l@{\hspace{1\tabcolsep}}l}
				Subgraph & \multicolumn{1}{c|}{A} & \multicolumn{2}{c|}{B} & \multicolumn{2}{c|}{C} & \multicolumn{2}{c}{D} \\
				& Simulation & Simulation & CPM Lab & Simulation & CPM Lab & Simulation & CPM Lab \\ \hline
				Reference error$^*$ [\si{\milli\meter}] & 40.1 (35.9) & 30.7 (28.9) & 27.9 (30.5) & 30.8 (28.8) & 19.9 (23.7) & 19.4 (17.6) & 17.0 (16.1) \\
				Time for one lap [\si{\second}] & 13.8 & 15.8 & 15.8 & 15.8 & 16.0 & 16.2 & 16.0 \\
				Max. expanded vertices & 159114 & 1966 & 430 & 1969 & 322 & 879 & 291 \\
				Computation time [\si{\milli\second}] & 9738.3 (5228.1) & 125.0 (67.5) & 12.4 (3.8) & 166.5 (80.1) & 11.8 (3.1) & 138.4 (71.2) & 10.2 (3.3) \\
				Max. comp. time [\si{\milli\second}] & 28436.7 & 600.9 & 32.3 & 570.7 & 24.5 & 654.4 & 32.8 \\
		\end{tabular} \newline \newline
		\normalsize{$^*$ Euclidean distances to the center of the lane.}
		\label{tab:results}
	\end{sidewaystable}

	\section{Conclusions}
	In this paper, we describe the construction of a motion primitive automaton with Pareto-optimal motion primitives.
	We evaluate four configurations of a generic automaton in both simulation and experiments in the \ac{cpmlab}, for multiple vehicles, in a prioritized \ac{dmpc} scheme. The results show that the choice of primitives
	and their optimization criteria play a role in the planning's real-time capability, trajectory performance, computation time, and speed of the vehicles, causing the need for a trade-off between these criteria. Future work can evaluate a reconfiguration of the automaton in real-time, based on a dynamic objective's prioritization.


\begin{thebibliography}{10}
		
		\bibitem{Fla2013}
		K.~Fla{\ss}kamp, {\em On the Optimal Control of Mechanical Systems -- Hybrid
			Control Strategies and Hybrid Dynamics}.
		\newblock PhD thesis, University of Paderborn, 2013.
		
		\bibitem{klock-cyber-lab-2021}
		M.~Kloock, P.~Scheffe, J.~Maczijewski, A.~Kampmann, A.~Mokhtarian,
		S.~Kowalewski, and B.~Alrifaee, ``{Cyber-Physical Mobility Lab}: An
		open-source platform for networked and autonomous vehicles,'' in {\em 2021
			European Control Conference (ECC)}, pp.~1937--1944, 2021.
		
		\bibitem{frazzoli2005}
		E.~Frazzoli, M.~A. Dahleh, and E.~Feron, ``Maneuver-based motion planning for
		nonlinear systems with symmetries,'' {\em IEEE Transactions on Robotics},
		vol.~21, no.~6, pp.~1077--1091, 2005.
		
		\bibitem{commonroad-Althoff2017a}
		M.~Althoff, M.~Koschi, and S.~Manzinger, ``Commonroad: Composable benchmarks
		for motion planning on roads,'' in {\em Proc. of the IEEE Intelligent
			Vehicles Symposium}, 2017.
		
		\bibitem{scheffe2022ieee}
		P.~Scheffe, M.~V.~A. Pedrosa, K.~Flaßkamp, and B.~Alrifaee, ``Receding horizon
		control using graph search for multi-agent trajectory planning,'' {\em IEEE
			Transactions on Control Systems Technology}, pp.~1--14, 2022.
		
		\bibitem{learning-pedrosa2022}
		M.~V.~A. Pedrosa, T.~Schneider, and K.~Flaßkamp, ``Learning motion primitives
		automata for autonomous driving applications,'' {\em Mathematical and
			Computational Applications}, vol.~27, no.~4, 2022.
		
		\bibitem{Flakamp2019SymmetryAM}
		K.~Fla{\ss}kamp, S.~Ober-Bl{\"o}baum, and K.~Worthmann, ``Symmetry and motion
		primitives in model predictive control,'' {\em Math. Control. Signals Syst.},
		vol.~31, pp.~455--485, 2019.
		
		\bibitem{pedrosa2021}
		M.~V.~A. Pedrosa, T.~Schneider, and K.~Flaßkamp, ``Graph-based motion planning
		with primitives in a continuous state space search,'' in {\em 2021 6th
			International Conference on Mechanical Engineering and Robotics Research
			(ICMERR)}, pp.~30--39, 2021.
		
		\bibitem{naumann2018_coincar}
		M.~Naumann, M.~Lauer, and C.~Stiller, ``Generating comfortable, safe and
		comprehensible trajectories for automated vehicles in mixed traffic,'' in
		{\em 2018 21st International Conference on Intelligent Transportation Systems
			(ITSC)}, pp.~575--582, 2018.
		
		\bibitem{scheffe2022feasibility}
		P.~Scheffe, G.~Dorndorf, and B.~Alrifaee, ``Increasing feasibility with dynamic
		priority assignment in distributed trajectory planning for road vehicles,''
		in {\em 2022 IEEE 25th International Conference on Intelligent Transportation
			Systems (ITSC)}, pp.~3873--3879, 2022.
		
		\bibitem{alrifaee2016}
		B.~Alrifaee, F.-J. He{\ss}eler, and D.~Abel, ``Coordinated non-cooperative
		distributed model predictive control for decoupled systems using graphs,''
		{\em IFAC-PapersOnLine}, vol.~49, no.~22, pp.~216--221, 2016.
		
		\bibitem{scheffe_muCars}
		P.~Scheffe, J.~Maczijewski, M.~Kloock, A.~Kampmann, A.~Derks, S.~Kowalewski,
		and B.~Alrifaee, ``Networked and autonomous model-scale vehicles for
		experiments in research and education,'' {\em IFAC-PapersOnLine}, vol.~53,
		no.~2, pp.~17332--17337, 2020.
		
	\end{thebibliography}
\end{document}